\documentclass[12pt]{iopart}

\usepackage[T1]{fontenc}
\usepackage{ae,aecompl}
\usepackage[normalem]{ulem}
\usepackage{graphicx}	
\expandafter\let\csname equation*\endcsname\relax
\expandafter\let\csname endequation*\endcsname\relax
\usepackage{amsmath, mathtools, xfrac}	
\usepackage{amssymb, enumitem}	
\usepackage{longtable, multicol}
\usepackage{rotate, setspace}
\usepackage{lscape}
\usepackage{natbib, hyperref}
\usepackage{latexsym,verbatim,xcolor}

\usepackage{url}

\begin{document}

\title[Poorly Studied King 2 and King 5]{Study of Open Star Clusters Using the Gaia DR3: I-Poorly Studied King 2 and King 5}

\author{A. A. Haroon$^{1,2}$, W. H. Elsanhoury$^{3,*}$, E. A. Elkholy$^{2,3}$, A. S. Saad$^{2,4}$ D. C. \c{C}{\i}nar$^5$}

\address{$^1$Astronomy and Space Science Department, Faculty of Science, King Abdulaziz University, Jeddah, Kingdom of Saudi Arabia}

\address{$^2$Astronomy Department, National Research Institute of Astronomy and Geophysics (NRIAG) 11421, Helwan, Cairo, Egypt}

\address{$^3$Physics Department, College of Science, Northern Border University, Arar, Saudi Arabia}

\address{$^4$Department of Management Information System, College of Business and Economics, Qassim University, P.O. BOX 6666, Buraidah 51452, Saudi Arabia}

\address{$^5$Istanbul University, Institute of Graduate Studies in Science, Programme of Astronomy and Space Sciences, 34116, Beyaz{\i}t, Istanbul, Turkey}

~~~~~~$^{*}$ Author to whom any correspondence should be addressed.\ead{elsanhoury@nbu.edu.sa}
\vspace{10pt}
\begin{indented}
\item[]XX XXXX (XX XX)
\end{indented}

\begin{abstract}
In this study, we utilize photometric and kinematic data from \textit{Gaia} DR3 and the {\sc ASteCA} package to analyze the sparsely studied open clusters, King 2 and King 5. For King 2, we identify 340 probable members with membership probabilities exceeding 50\%. Its mean proper motion components are determined as $(\mu_\alpha\cos\delta,~\mu_\delta) = (-1.407 \pm 0.008, -0.863 \pm 0.012)$ mas yr$^{-1}$, and its limiting radius is derived as 6.94$_{-1.06}^{+0.22}$ arcminutes based on radial density profiles. The cluster has an estimated age of $4.80 \pm 0.30$ Gyr, a distance of $6586 \pm 164$ pc, and a metallicity of [Fe/H] = -0.25 dex ($z = 0.0088$). We detect 17 blue straggler stars (BSSs) concentrated in its core, and its total mass is estimated to be $356 \pm19~M_{\odot}$. The computed apex motion is $(A_o,~D_o) = (-142^\circ_{.}61 \pm 0^\circ_{.}08, -63^\circ_{.}58 \pm0^\circ_{.}13)$. Similarly, King 5 consists of 403 probable members with a mean proper motion components $(\mu_\alpha\cos\delta,~\mu_\delta) = (-0.291 \pm 0.005, -1.256 \pm 0.005)$ mas yr$^{-1}$ and a limiting radius of 11.33$_{-2.16}^{+5.45}$ arcminutes. The cluster's age is determined as $1.45 \pm 0.10$ Gyr, with a distance of $2220 \pm 40$ pc and a metallicity of [Fe/H] = -0.15 dex ($z = 0.0109$). We identify 4 centrally concentrated BSSs, and the total mass is estimated as $484 \pm 22~M_{\odot}$. The apex motion is calculated as $(A_o,~D_o) = (-115^\circ_{.}10 \pm 0^\circ_{.}09, -73^\circ_{.}16 \pm 0^\circ_{.}12)$. The orbital analysis of King 2 and King 5 indicates nearly circular orbits, characterized by low eccentricities and minimal variation in their apogalactic and perigalactic distances. King 2 and King 5 reach maximum heights of \(499 \pm 25\) pc and \(177 \pm 2\) pc from the Galactic plane, respectively, confirming their classification as young stellar disc population.
\end{abstract}

%
\vspace{2pc}

\section{Introduction}\label{sec1}
Open clusters (OCs) are fundamental probes for investigating the formation and evolution of the Milky Way. Their broad age distribution offers a timeline for studying the development of the Galactic disc, from its early formation to more advanced stages. The spatial distribution and kinematic properties of OCs provide important insights into the gravitational potential and the perturbations affecting the structure and dynamics of the Galaxy. By examining the evolutionary paths and eventual dissolution of OCs, we can better understand the processes behind the assembly and long-term evolution of the Galactic disc, as well as the formation and structure of spiral galaxies. These analyses are essential for building a comprehensive model of Galactic evolution. Most OCs dissipate completely within a few million years \citep{SpitzerHart1971, Bhattacharya2022, Liu2025}. The OCs known to be older than 1 Gyr are thought to have persisted because of their specific orbital characteristics, which help them to avoid close encounters with the Galactic plane \citep{Friel1995}. Several dynamical processes contribute to the eventual disruption of OCs, including internal interactions between member stars, stellar evolution, collisions with giant molecular clouds, and gravitational influences from the Galactic potential \citep[see e.g.,][]{Gustafsson2016}. 

OCs play an important role in the calibration of the cosmic distance scales due to the precise determination of their distances. Their well-constrained ages and metallicities make them key benchmarks for stellar evolution models and standard candles \citep[e.g.,][]{Gaia2018, Bossini2019, Cantat2020, Gaia2021, Gaia2023} On the other hand, to constrain the initial luminosity functions (LF) and initial mass functions (IMF) in stellar populations. In addition, the radial velocities of OCs have been used to study local Galactic kinematics, including the motion of the Sun, the Oort constants ($A~\&~B$), and the rotation curve of the Galaxy. Older clusters, especially those at larger distances, provide important data for identifying metallicity gradients within the Galactic disc, relating cluster age to metallicity, and revealing the complex history of chemical enrichment and mixing processes within the disc \citep{Friel1995, Alfonso2024}.

The dynamics of OCs, which are intimately linked to their orbital properties and metallicity, provide deep insights into the processes governing their formation and evolution within the Galactic disc. As clusters pass through the disc, their orbits, shaped by gravitational interactions and the surrounding stellar populations, provide clues to their past trajectories and the environmental conditions that have influenced their evolution. At the same time, the metallicity of these clusters serves as a key indicator of their formation history, reflecting the chemical evolution of the disc and the enrichment processes that have taken place over cosmic timescales \citep{Nilakshi2002, Ishchenko2025}. This intricate relationship between orbital dynamics and metallicity not only enhances our understanding of individual clusters, but also contributes to a more holistic understanding of Galactic structure, stellar population dynamics, and the broader implications for stellar evolution in different environments.

\subsection{Previous Studies of King 2 and King 5}

King 2 is among the oldest open clusters (OC) in the Milky Way, with an estimated age of 6 Gyr and a distance of 5700 pc (Table \ref{tab:table_01}). It has not been extensively studied due to its substantial distance and lack of detailed membership data. The coordinates of King 2 are ($\alpha,~\delta)\textsubscript{2000} = (00^{\rm h} 51^{\rm m} 00^{\rm s}, +58^{\rm d} 11^{\rm m} 00^{\rm s}$) and ($l,~b$) = ($122^{\circ}$.874, -4$^{\circ}$.688). The first optical CMD for this distant cluster was presented by \citet{Kaluzny1989b} using $BV$ CCD photometry, giving a range of possible ages and distances based on various assumptions about reddening and metallicity. The distance of King 2 from the Galactic center was estimated to be about 14 kpc. A more comprehensive study was subsequently carried out by \citet[][A90 hereafter]{Aparicio1990} using $UBVR$ photometric data. Assuming solar metallicity, they determined the age of the cluster to be 6 Gyr and the distance to be 5.7 kpc. They also reported significant binarity in the main sequence (MS) population.

Using spectroscopic data, \citet[][WC09 hereafter]{WarrenCole2009} determined a metallicity of [Fe/H] = -0.42 $\pm$ 0.09 (dex). This iron abundance is poorer than the estimated solar abundance and contradicts the findings of A90. WC09 also finds a distance of 6.5 kpc and a slightly younger age of about 4 Gyr, with an CMD for the 2MASS $K_s$ cluster reddening $E{(B-V)} = 0.31$ mag, which gives a better fit if a reddening value of $0.31$ mag is adopted. This distance places King 2 at a Galactic radius of about 13 kpc, and its metallicity is in good agreement with the trends observed in the Galactic abundance gradients derived by \cite{Joshi2024}. Until the publication of $Gaia$ DR2 \citep{Gaia2018}, no proper motion studies had been performed for this cluster. However, \cite{Cantat2018} provided a membership catalogue for King 2, identifying 128 members using $Gaia$ DR2 data. Subsequently, \cite{Jadhav2021} presented a kinematic membership analysis using $Gaia$ DR3 kinematic data, identifying 1072 stars, of which 340 are classified as possible members. The optical photometric studies mentioned above indicate a significant presence of post-main-sequence hot stars in King 2. 

\cite{PhelpsJanesMontgomery1994} and \cite{CarraroVallenari2000} studied King 5 photometrically and found it to be of Hyades age ($\sim 1$ Gyr old). \cite{CarraroVallenari2000} estimated the distance of King 5 located at ($\alpha,~\delta$)\textsubscript{2000} = ($03^{\rm h}~14^{\rm m}~45^{\rm s},~52^{\rm d}~41^{\rm m}~12^{\rm s}$) and ($l,~b$) = (143$^{\circ}$.776, -4$^{\circ}$.287) as 1.90 kpc from the Sun. \cite{DurgapalPandeyMohan1998} presented preliminary findings on this cluster. 

The analysis of various kinematic parameters enhances our understanding of the dynamics of these OCs, establishing a framework for future investigations. Our findings underscore the relevance of these OCs in the context of stellar population studies, paving the way for further research into their formation and evolution. The classification of King 2 and King 5 within the young disk of the Milky Way reflects their relative youth compared to older globular clusters, which exceed ten billion years in age. Despite being over a billion years old, these clusters reside in a dynamic region characterized by ongoing star formation and evolving stellar populations. Their inclusion in the young disk highlights the complex interplay of age, kinematics, and chemical characteristics that define stellar systems in this region. This perspective is supported by studies on the young disk's formation history and dynamics, emphasizing the importance of recognizing these older OCs as part of a more active stellar environment.

The article is divided into the following sections: Section \ref{sec2} deals with a description of the astrometric and photometric data from the $Gaia$ DR3 catalog. Section \ref{sec3} is devoted to the structural properties and derivation of their fundamental parameters. In Section \ref{sec4}, we present our study of their photometric parameters (reddening, distance modulus, and ages) while the luminosity and mass functions are described in detail in Section \ref{sec5}. In Section \ref{sec6}, we discuss their dynamic evolution at different times. Section \ref{sec7} presents velocity ellipsoid parameters, convergent point analysis, and their morphology in 3D. Our conclusions are given in Section \ref{sec8}.
\begin{table}
\small
\centering
\caption{The estimated values of color excess ($E(B-V)$), distance ($d$), metallicity ([Fe/H]) and age ($t$) for King 2 and King 5.}
\label{tab:table_01}
\begin{tabular}{ccccc}
\hline
$E(B-V)$     & $d$       & [Fe/H]      & $t$          & Ref \\
(mag)        & (pc)    & (dex)         & (Gyr)        &     \\
\hline
\hline
             &         & King 2        &              &     \\
\hline
0.23 to 0.50 & 7000    & --            & 4 to 6 & (1)   \\
0.31 $\pm$ 0.02& 5697 $\pm$ 65 & -0.50 to -2.2 & 6.0      & (2)   \\
0.31         & 5750    & -0.42 $\pm$ 0.09         & 6.02         & (3)   \\
\hline
\hline
             &         & King 5        &              &     \\
             \hline
--           & --      & --            & 1         & (4)   \\
--           & 1900    & 0.08          & 1         & (5) \\ 
\hline
\end{tabular}
\\
\noindent{
(1)~\citet{Kaluzny1989b}, (2)~\citet{Aparicio1990}, (3)~\citet{Dias2002}, (4)~\citet{PhelpsJanesMontgomery1994}, (5)~\citet{CarraroVallenari2000}.
}
\end{table}%

\section{Data and Analysis}\label{sec2}
\subsection{The Gaia DR3 Data}
\begin{figure}
\centering
\includegraphics[width=0.8\linewidth]{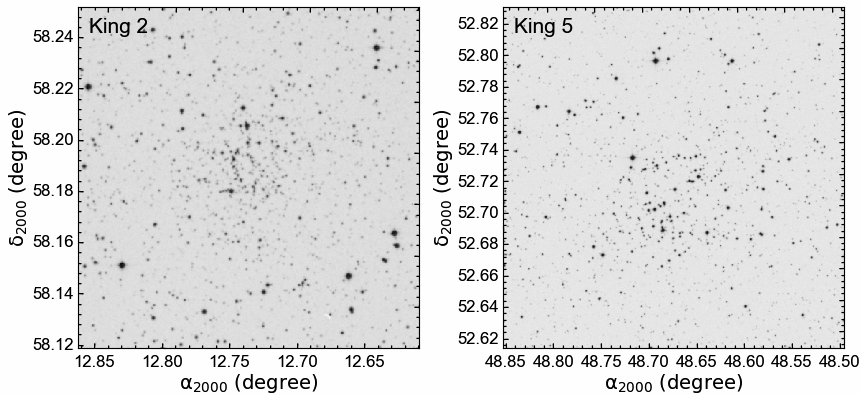}
\caption{Identiﬁcation maps of both OCs; King 2 (left) and King 5 (right).}
\label{fig:figure_01}
\end{figure}

We used fundamental parameters from prior studies on the open clusters King 2 and King 5 as initial values for this analysis. For the purpose of our analysis of the two clusters, we have downloaded the astrometric and photometric data with a radius of 30$'$ from the $Gaia$ Data Release \citep{Gaia2023}\footnote{\url{https://cdsarc.cds.unistra.fr/viz-bin/cat/I/355}} hereafter DR3. Identification maps of King 2 and King 5 are shown in Figure \ref{fig:figure_01}. $Gaia$ provides raw data collected by the European Space Agency's (ESA) $Gaia$ mission\footnote{https://www.cosmos.esa.int/gaia} during its continuous sky scanning by the $Gaia$ Data Processing and Analysis Consortium (DPAC)\footnote{https://www.cosmos.esa.int/web/gaia/dpac/consortium}. $Gaia$ DR3 represents a significant advance in the mission's objectives for stellar, galactic and extragalactic studies. This release indicates that the evolution of the catalogue has become significantly more stable, exhibiting far fewer alterations between $Gaia$ DR2 and DR3 than observed during the transition from $Gaia$ DR1 to DR2. Notably, 97.5\% of the catalogued sources have been maintained consistently across these two releases. The updated merging and split criteria reduce the number of superseded as well as new sources. $Gaia$ DR3 has revolutionized the study and analysis of OC morphology by providing parallaxes with $\sim30\%$ higher precision and proper motions with double accuracy, as compared to those in the $Gaia$ DR2 \citep{Gaia2018}.

$Gaia$ DR3 provides five key quantities for nearly 1.8 billion stars and non-stellar objects \citep{Gaia2023}, including their central coordinates ($\alpha,~\delta$), proper motion components ($\mu_\alpha\cos\delta,~\mu_\delta$), and trigonometric parallaxes ($\varpi$). The dataset covers sources with a limiting magnitude range from 3 to 21 in the $G$ band and includes radial velocity measurements ($V_{\rm r}$) for a limited number of entries. The precision of the $Gaia$ DR3 astrometry is exceptional, with parallax errors ranging from 10 to 100 $\mu$as and proper motion uncertainties generally being more precise between 20 and 140 $\mu$as yr$^{-1}$ though it is acknowledged that some proper motion errors may be negative. In addition to the astrometric data, $Gaia$ photometry is provided in three bands: $G$, $G_{\rm BP}$, and $G_{\rm RP}$ \citep{Riello2021}, covering the optical spectrum from 330 to 1050 nm, 330 to 680 nm, and 630 to 1050 nm, respectively. To enhance data quality, $Gaia$ has improved its methods for estimating background noise in all three photometric bands. This ensures more accurate measurements of celestial objects by accounting for background light variations.

\begin{figure}
\centering
\includegraphics[width=0.99\linewidth]{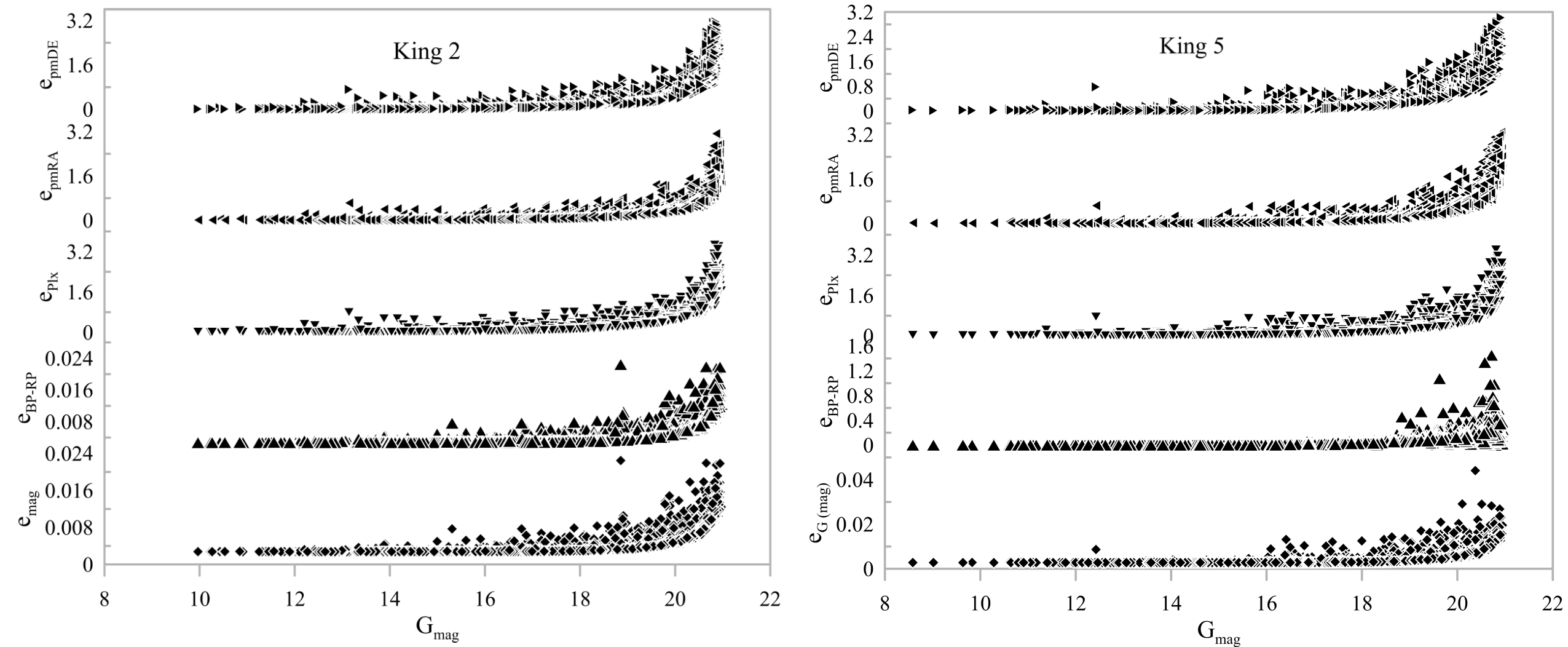}
\caption{The \textit{left panel} and \textit{right panel} are the photometric errors in $Gaia$ $G$ errors, color index ($G_{\rm BP} - G_{\rm RP}$), parallax, and proper motions versus $G$ magnitude passbands for King 2 and King 5 respectively.}
\label{fig:figure_02}
\end{figure}

In $Gaia$ DR3 the number of available observations per source is on average higher than in $Gaia$ DR2. The parallax uncertainties for $G\leq$ 15 mag sources are typically in the range of 0.02 - 0.03 mas, and about 0.07 mas for $G\sim$17 mag sources. For the proper motion components, the uncertainties are as low as 0.01 - 0.02 mas yr$^{-1}$ for sources with $G<$ 15 mag, increasing to approximately 0.4 mas yr$^{-1}$ for sources with $G\sim$ 20 mag. Table \ref{tab:table_new_rev1} shows the mean errors for $G$ magnitudes and $(G_{\rm BP}-G_{\rm RP})$ color indices of stars in the King 2 and King 5 clusters as a function of $G$ magnitudes. Figure \ref{fig:figure_02}, left and right panels, shows the photometric errors in $G$ band, the color index ($G_{\rm BP}-G_{\rm RP}$), the trigonometric parallax and the proper motion components as a function of $G$ magnitude for King 2 and King 5 for sources within 30', respectively, based on $Gaia$ DR3 data.

\begin{table}
\centering
\label{tab:table_new_rev1}
\small
\caption{Mean internal photometric errors for $G$ and $(G_{\rm BP}-G_{\rm RP})$ magnitudes per $G$ magnitude bin for the King 2 and King 5 OCs for sources within 30'.}
\begin{tabular}{c|ccc|ccc}
 \hline
 &  &    King 2    &          &  &    King 5    &           \\
 \hline
 $G$& $N$ & $\sigma_{\rm G}$ & $\sigma_{G_{\rm BP}-G_{\rm RP}}$ & $N$ & $\sigma_{\rm G}$ & $\sigma_{G_{\rm BP}-G_{\rm RP}}$  \\
 \hline
(06, 14]  & 183    & 0.0028 & 0.0028   & 171    & 0.0028 & 0.0024    \\
(14, 15]  & 218    & 0.0028 & 0.0028   & 212    & 0.0028 & 0.0021    \\
(15, 16]  & 462    & 0.0028 & 0.0028   & 429    & 0.0028 & 0.0021    \\
(16, 17]  & 854    & 0.0028 & 0.0028   & 797    & 0.0029 & 0.0058    \\
(17, 18]  & 1550   & 0.0029 & 0.0029   & 1416   & 0.0030  & 0.0152    \\
(18, 19]  & 2402   & 0.0032 & 0.0032   & 2280   & 0.0033 & 0.0373    \\
(19, 20]  & 3259   & 0.0040  & 0.0040    & 3549   & 0.0045 & 0.0818    \\
(20, 21]  & 3309   & 0.0071 & 0.0071   & 3612   & 0.0083 & 0.1526 \\  
 \hline
\end{tabular}
\end{table}

\subsection{Data Analysis}
A new set of open-source tools has been developed for the study of structural parameters (e.g., radial density profiles, clusters radii, and membership probabilities), as well as cluster’s fundamental parameters (reddening, distance modulus, age, and mass) either automatically or semi-automatically with the aid of the Automated Stellar Cluster Analysis package ({\sc ASteCA}; \cite{Perren2015})\footnote{\url{https://asteca.readthedocs.io/en/latest/about.html}}. A detailed description of the functions built within this tool can be found in \cite{Perren2015}. The {\sc ASteCA} code is comprised of three principal independent analysis modules. The initial module is the structure study, which entails the identification of a cluster region based on an observed over-density of stars. The second module is concerned with the estimation of individual membership probabilities for stars situated within the over-density region. The final module is dedicated to the search for the best-fit parameters that describe the cluster's fundamental properties.

\section{Structural Parameters}\label{sec3}
\subsection{Determination of the Clusters' Centers}
\begin{figure}
\centering
\includegraphics[width=15cm,height=10cm]{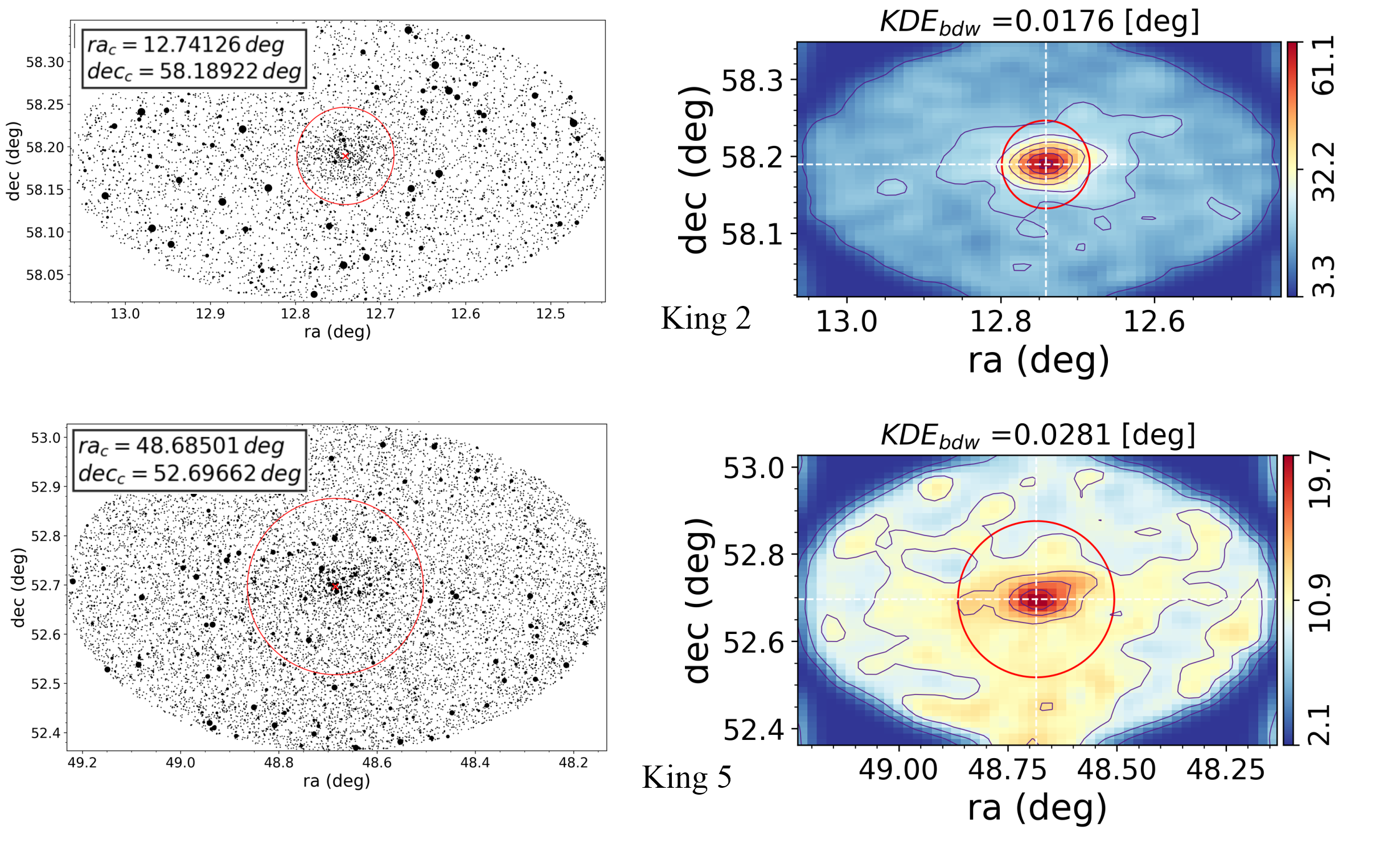}
\caption{Left panel: the cluster's map charts defining the cluster regions with red circle centered by x-red symbol. Right panel: the contours of the cluster's surface density (stars per arcmin$^{-2}$) with color bars ranged from low (blue) to high (red) for both King 2 and King 5 OCs in the upper and lower panels, respectively.}
\label{fig:figure_03}
\end{figure}

The central coordinates of King 2 and King 5 were estimated using the {\sc ASteCA} package, which adheres to the established methodology of assigning the clusters' center to the location of maximum spatial density \citep{Perren2015}, {\sc ASteCA} achieves this by fitting a two-dimensional Gaussian kernel density estimator (KDE) to the positional diagram of the cluster and identifying the peak value, as illustrated in Figure \ref{fig:figure_03}. In contrast to other algorithms that necessitate initial input values, {\sc ASteCA} operates without them in fully automatic mode (although semi-automatic operation is feasible), thereby ensuring consistent convergence.

Furthermore, this method eliminates the necessity for region binning, as the kernel bandwidth is calculated using Scott’s rule \citep{Scott1992}. By concurrently estimating the maximum density in both spatial dimensions, it circumvents potential inaccuracies in central coordinate determination, particularly in densely populated fields. Additionally, this approach is independent of the coordinate system, rendering it suitable for positional data stored in either pixel units or degrees. We redetermined the centers in right ascension and declination for King 2 and King 5 using {\sc ASteCA} which is shown in Figure \ref{fig:figure_03}. The contours indicates cluster surface density (stars per arcmin$^{-2}$), where the blue and red colors refers to low and high densities regions for King 2 and King 5, respectively.

For King 2, equatorial coordinates ($\alpha, \delta$) were determined by this study as $(00^{\rm h}50^{\rm m}57{^{\rm s}}.90$, $58^\circ 11'21".20$), compared to ($00^{\rm h}50^{\rm m}57{^{\rm s}}.30$, $58^\circ 11'14".49$) in \citet{Hunt2023}. Similarly, the Galactic coordinates of the OCs ($l$, $b$) were recalculated as ($123^\circ.26$, $-04^\circ.41$), in this study, while \cite{Hunt2023} found it to be ($122^\circ.868$, $-04^\circ.686$). For King 5, the equatorial coordinates ($\alpha, \delta$) were determined to be ($03^{\rm h} 14^{\rm m} 44{^{\rm s}}.40$, $52^\circ 41' 47{^{\rm "}}.83$) in this study, compared to ($03^{\rm h} 14^{\rm m} 42{^{\rm s}}.53$, $52^\circ42'11".28$) in \cite{Hunt2023}. The Galactic coordinates ($l$, $b$) were determined as $(144^\circ.14$, $-03^\circ.83$) in this study and $(143^\circ.761$, $-04^\circ.276$) in \cite{Hunt2023}, slightly different from the earlier report. This comparison demonstrates that, while the recalculated center positions for both clusters show slight deviations from the values reported by \cite{Hunt2023}, these differences are within acceptable margins of uncertainty, suggesting that the new estimations refine rather than significantly alter the previously established coordinates, thereby enhancing the precision of the cluster's spatial parameters.

\subsection{Radial Density Profiles and Radius of the Clusters}
\begin{figure}
\centering
\includegraphics[width=15cm,height=10cm]{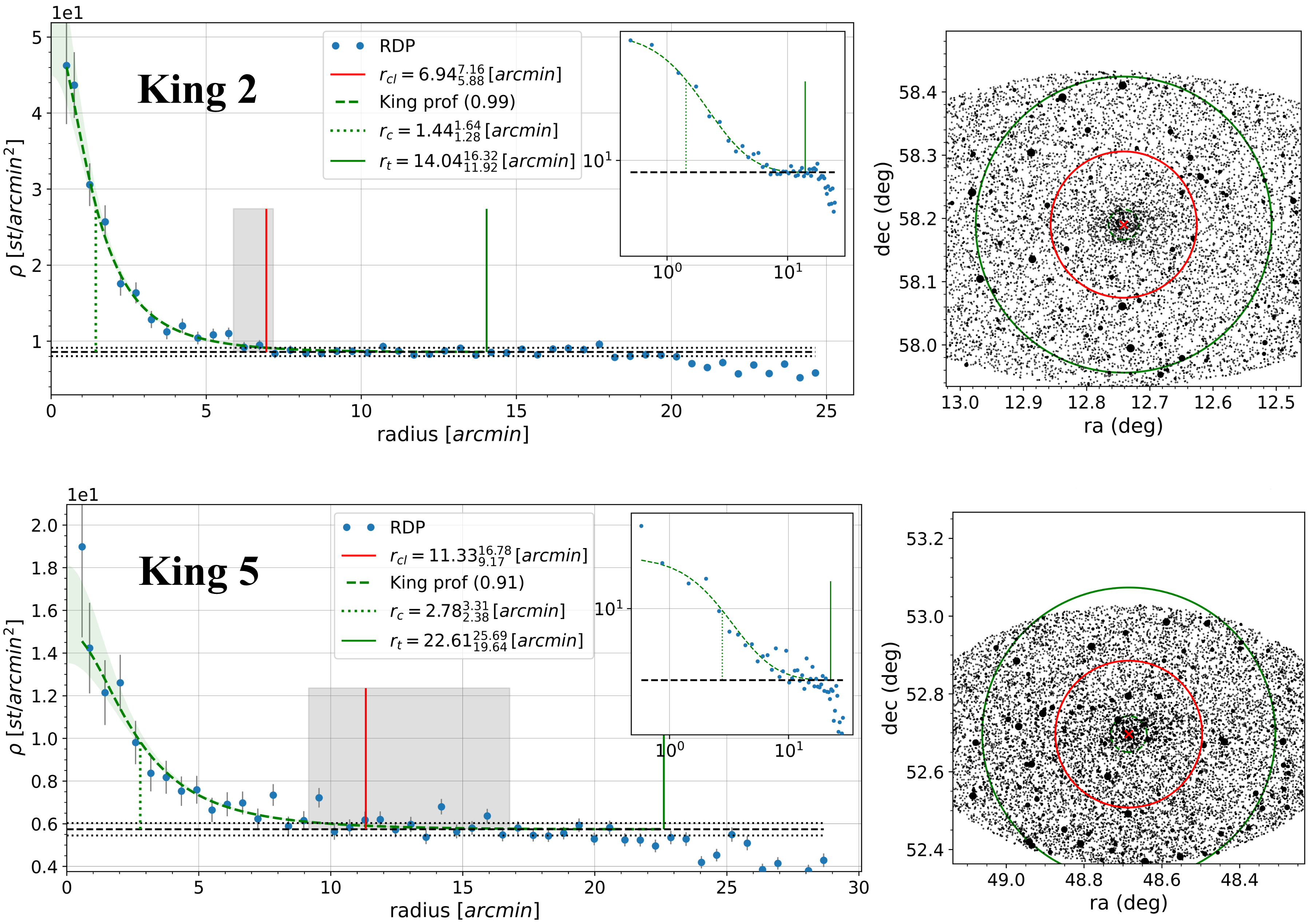}
\caption{The RDP for King 2 and King 5 OCs were generated using the {\sc ASteCA} code, with the results shown as blue dots. The green dashed line and shaded region represent King’s density profile. Additionally, the black dashed line indicates the background field density ($\rho_{\rm bg}$), while the dotted black line marks the central surface density ($\rho_{\rm o}$). Vertical lines are used to denote the structural parameters, including the core radius ($r_{\rm c}$), the limiting radius ($r_{\rm cl}$), and the tidal radius ($r_{\rm t}$).}
\label{fig:figure_04}
\end{figure}

After we re-estimated the centers, the next step was to construct a radial density profile (RDP); every point of the RDP was obtained by generating rings ($i^{th}$ zones) around the center defined for the potential cluster, i.e., the comparison field. The number density of stars ($\rho_i$) in the $i^{th}$ zone is calculated as $\rho_i=N_i/A_i$, where $N_i$ and $A_i$ denote the number of stars and the area of the $i^{th}$ zone, respectively. Creating a King profile \citep{King1962} for this distribution, a smooth dashed line has been produced with the assistance of the {\sc ASteCA} code \citep{Perren2015}. The method allowed us to estimate the internal cluster structural parameters as shown in Figure \ref{fig:figure_04}, where the values have large uncertainties, this might be due to the non-spherical geometry of these OC regions. The King's density ($\rho(r)$) profile \citep{King1962}, depends on core radius ($r_c$) (i.e., radial distance) at which the value of $\rho(r)$ becomes half of the central density ($\rho_o$) and on the background surface density ($\rho_{bg}$). In expansion, we will define the cluster radius ($r_{\rm cl}$) at which the King profile intersects with background surface density $\rho_{bg}$. At this point the background star density $\rho_b$ is given by ($\rho_b=\rho_{bg}+3\sigma_{\rm bg}$), where $\sigma_{\rm bg}$ is the uncertainty of $\rho_{bg}$ \citep{Bukowiecki2011}.

In this study, we have calculated both the density contrast parameter ($\delta_{\rm c}$) and the concentration parameter ($C$) for these clusters for the first time.$\delta_{\rm c}$ is defined as the stellar density contrast of the clusters in question relative to the background population (i.e., $\delta_{\rm c} = 1 + \rho_{\rm o}/\rho_{\rm bg}$), therefore, its a measure of the compactness of the cluster \citep{BonattoBica2009} and $C$ is defined by \cite{PetersonKing1975} represents the ratio between $r_{cl}$ and $r_c$, thereby providing insight into the cluster's structure. The computed values for both the density contrast parameter and the concentration parameter for King 2 and King 5 are presented in Table \ref{tab:table_02}.

It is notable that $\delta_c$ assumes relatively high values, spanning the range $7\lesssim\delta_c\lesssim23$ \citep{BonattoBica2009}. We calculated $\delta_c$ as 6.64 $\pm$ 0.44 and 3.66 $\pm$ 0.68 for the clusters King 2 and King 5, respectively. Since these two clusters are old ones, as the ages of OC’s increase, stellar dynamics within the clusters influence the central part of the cluster to become circular \citep{Fujii2016, Qin2024, Pera2024}.

\begin{table}
\renewcommand{\arraystretch}{1.5}
\centering
\tiny
\caption{Structural properties of the King 2 and King 5 OCs obtained in this study.}
\label{tab:table_02}
\begin{tabular}{ccc|ccc|cc|cc}
\hline
Cluster	& $\rho_{\rm o}$&$\rho_{\rm bg}$&$r_{\rm cl}$&$r_{\rm c}$&$r_{\rm t}$&$r_{\rm c}$&$r_{\rm t}$&	$\delta_{\rm c}$&$C$\\
& \multicolumn{2}{c}{(stars arcmin$^{-2}$)}& \multicolumn{3}{c}{(arcmin)\textsubscript{ASteCA}}& \multicolumn{2}{c}{(pc)}&&\\
\hline
King 2& 45.90$\pm$0.67	& 8.15$\pm$1.60	& ~6.94$_{-1.06}^{+0.22}$
& 1.44$_{-0.16}^{+0.20}$& 14.04$_{-2.12}^{+2.28}$& 3.15$\pm$0.56	&30.83$\pm$5.55& 6.64$\pm$0.44& 4.83$\pm$0.46 \\
King 5& 15.30$\pm$0.39	& 5.75$\pm$3.45	& 11.33$_{-2.16}^{+5.45}$& 2.78$_{-0.40}^{+0.53}$& 22.61$_{-2.97}^{+3.08}$& 2.04$\pm$0.70  &16.59$\pm$4.08& 3.66$\pm$0.68& 4.08$\pm$0.50 \\
\hline    
\end{tabular}
\end{table}

By combining both the $r_{\rm cl}$ and $r_{\rm c}$ to characterise clusters and their structures, \cite{PetersonKing1975} provided a framework for interpreting cluster sizes. \cite{Nilakshi2002} found that the angular size of the coronal region typically extends to about 6$r_{\rm c}$, while \cite{Bukowiecki2011} showed that $r_{\rm cl}$ can range from 2$r_{\rm c}$ to 7$r_{\rm c}$. In our analysis, as shown in Table \ref{tab:table_02}, $C$ for King 2 and King 5 are approximately 4.83 $\pm$ 0.46 and 4.08 $\pm$ 0.50, respectively, which agrees well with the results of \cite{Bukowiecki2011}. In addition, \cite{Hunt2023} reported the $r_{\rm c}$ for King 2 and King 5 to be 2.375 and 1.728 pc respectively, with $r_{\rm t}$ of 19.911 and 7.906 arcminutes, indicating that both clusters exhibit a well-defined structure and size that aligns with their dynamical status and evolutionary history as observed in previous studies.

\section{Photometric Analysis: Reddening, Distance and Age}
\label{sec4}
\subsection{Decontamination of the Cluster from Field Stars}
To determine the fundamental astrophysical properties of the clusters, including reddening, distance modulus, and age, a photometric analysis was conducted through the construction of a color-magnitude diagram (CMD). Identifying probable members is crucial in reducing contamination from field stars. Various decontamination techniques have been developed and extensively discussed in the literature to statistically eliminate field stars. Among these methods, proper motions (PMs) provide a robust approach for estimating membership probabilities with higher accuracy. In this study, membership probabilities were assigned based on the rigorous selection criteria implemented in the second module of the {\sc ASteCA} code, following the methodology described by \cite{CabreraCanoAlfaro1990, Elaziz2016}.

The study by \cite{Perren2015} utilized the {\sc ASteCA} code to estimate the number of probable cluster members through two different approaches. The first approach computes the radial density profile (RDP) within the tidal radius ($r_{\rm t}$), relative to the density of background stars, utilizing a King model with three parameters (3P). The effectiveness of this approach depends on obtaining a well-constrained $r_{\rm t}$ and achieving convergence in the 3P fitting process; otherwise, the estimated member count may be significantly overestimated. The second method relies on direct star counts, where the expected number of field stars ($n_{\rm fl}$) within the cluster area is calculated as the product of the background density ($d_{\rm field}$) and the cluster area ($A_{\rm cl}$), determined from the cluster radius ($r_{\rm cl}$). Subtracting this field star estimate from the observed total star count within $r_{\rm cl}$ ($n_{\rm cl+fl}$) yields the final estimated cluster membership ($n_{\rm cl}$).

Both methods depend on completeness limits, as they estimate the number of member stars down to the faintest observed magnitudes. Using the {\sc ASteCA} code, the most probable members of King 2 and King 5 clusters were determined as 340 and 403, respectively, with a membership probability threshold of $P_{\rm ASteCA} \geq 50\%$.

\begin{figure}
\centering
\includegraphics[width=0.75\linewidth]{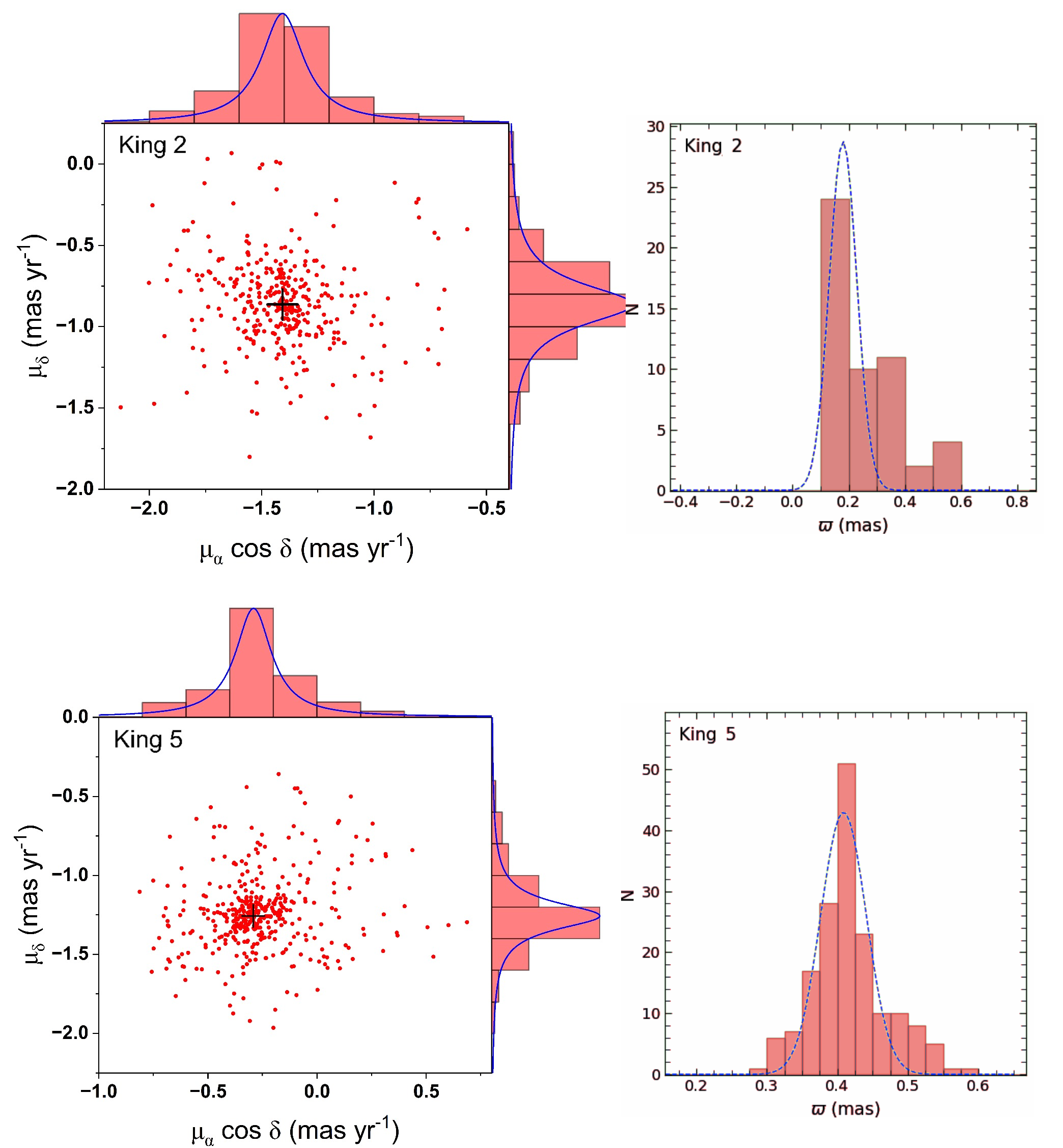}
\caption{\textit{Upper panel}: The distribution of the mean PM ($\mu_\alpha\cos\delta,~\mu_\delta$). \textit{Lower panel}: Normalised trigonometric parallax ($\varpi$) distribution for the entirety of the stellar member candidates within the cluster space}
\label{fig:figure_05}
\end{figure}

To derive the mean proper motion components of the clusters, Gaussian fitting was applied to the distributions of probable members in each coordinate direction, as shown in Figure \ref{fig:figure_05} (upper panel). To ensure the reliability of these results, the analysis was restricted to stars with positive parallax values, as depicted in Figure \ref{fig:figure_05}. The mean proper motion components in right ascension and declination were calculated for both clusters and are presented in Table \ref{tab:table_02}. Additionally, the trigonometric parallaxes of the member candidates were used to construct histograms for King 2 and King 5, from which the mean parallax values were obtained. These results indicate distances of $d_{\rm \varpi} = 5556 \pm 1512$ pc for King 2 and $d_{\rm \varpi} = 2439 \pm 190$ pc for King 5.

The {\sc ASteCA} code was further employed to estimate astrophysical parameters such as reddening, distance modulus, and age using photometric data in the $G$, $G_{\rm BP}$, and $G_{\rm RP}$ passbands. The code utilizes the PARSEC v1.2S\footnote{http://stev.oapd.inaf.it/cgi-bin/cmd} theoretical isochrones \citep{Bressan2012} and the \cite{Kroupa2002} initial mass function (IMF). As illustrated in Figure \ref{fig:figure_06}, the CMDs in $G$ vs. $(G_{\rm BP}-G_{\rm RP})$ were fitted with a dense grid of isochrones with metallicities $z = 0.0088$ and $z = 0.0109$, yielding ages of $t = 4.8 \pm 0.3$ Gyr for King 2 and $t = 1.45 \pm 0.1$ Gyr for King 5.

To convert the metallicity parameter $z$ into [Fe/H], the transformation provided by Bovy\footnote{https://github.com/jobovy/isodist/blob/master/isodist/Isochrone.py} was used, which is consistent with PARSEC isochrone models \citep{Yontan2022, Cakmak2024}:
\begin{equation}
z_{\rm x} = \frac{z}{0.7515 - 2.78 \times z}
\end{equation}
and
\begin{equation}
{\rm [Fe/H]} = \log \left(z_{\rm x} \right) - \log \left( \frac{z_{\odot}}{1 - 0.248 - 2.78 \times z_{\odot}} \right)
\end{equation}
where $z_{\rm x}$ is an intermediate variable, and the solar metallicity is assumed to be $z_{\odot} = 0.0152$ \citep{Bressan2012}. Applying these equations, the metallicity values for King 2 and King 5 were determined as $z=0.0088$ ([Fe/H] = $-0.25$ dex) and $z=0.0109$ ([Fe/H] = $-0.15$ dex), respectively.

To correct for the effects of interstellar extinction, the reddening values were adjusted using the relation $E(G_{\rm BP}-G_{\rm RP}) = 1.289 \times E(B-V)$ \citep{CasagrandeVandenBerg2018, Zhong2019}, while the extinction coefficient in the $G$ band was computed as $A_{\rm G} = 2.74 \times E(B-V)$. The estimated reddening values were verified using the Stilism 3D dust maps\footnote{https://stilism.obspm.fr} \citep{Capitanio2017}. The distance modulus derived from CMD fitting for King 2 and King 5 clusters was found to be $(m-M){\rm o} = 14.896 \pm 0.053$ and $13.321 \pm 0.038$ mag, respectively. These correspond to photometric distances of $d{\rm phot} = 6586 \pm 164$ pc for King 2 and $d_{\rm phot} = 2220 \pm 40$ pc for King 5. A summary of the obtained astrophysical and photometric parameters is provided in Table \ref{tab:table_03}.

Using our estimated distances $d$, we infer the clusters' distances to the Galactic center $R_{\rm gc}$ (i.e., $R_{\rm gc}=\sqrt{R_{\rm o}^{2}+(d\cos{b})^{2}-2R_{\rm o}d\cos{b}\cos{l}}$), where $R_{\rm o}=8.2 \pm 0.1$ kpc \citep{BlandHawthorn2019}, $l$ is the Galactic longitude, and $b$ is the Galactic latitude. The projected distances toward the Galactic plane ($X_{\odot}$, $Y_{\odot}$) and the distance from the Galactic plane ($Z_{\odot}$) takes the forms, $X_\odot=d\cos{b}\cos{l},~Y_\odot=d\cos{b}\sin{l}$, and $Z_\odot=d\sin{b}$. Results of these calculations for King 2 and King 5 are provided in Table \ref{tab:table_03}.

\subsection{Blue Straggler Stars}

The identification of blue straggler stars (BSSs) lying above the MS turnoff region in CMDs depends on detailed analysis of photometric data, in particular precise measurements from {\it Gaia} DR3, which provides valuable insights into the membership probabilities and cluster characteristics \citep{Leiner2021, Jadhav2021b, Chand2024}. BSS formation mechanisms, supported by models from \citet{Antonini2016} and \citet{Sindhu2019}, suggest that dense cluster environments provide ideal conditions for the formation of these massive stars, either through close stellar encounters or within primordial binaries. \citet{Gosnell2015} have shown that in older OCs, a significant fraction of BSS populations are likely to evolve from binary stars interactions - a formation pathway that may also be relevant for King 2 and King 5, given their ages.

\begin{figure}
\centering
\includegraphics[width=0.75\linewidth]{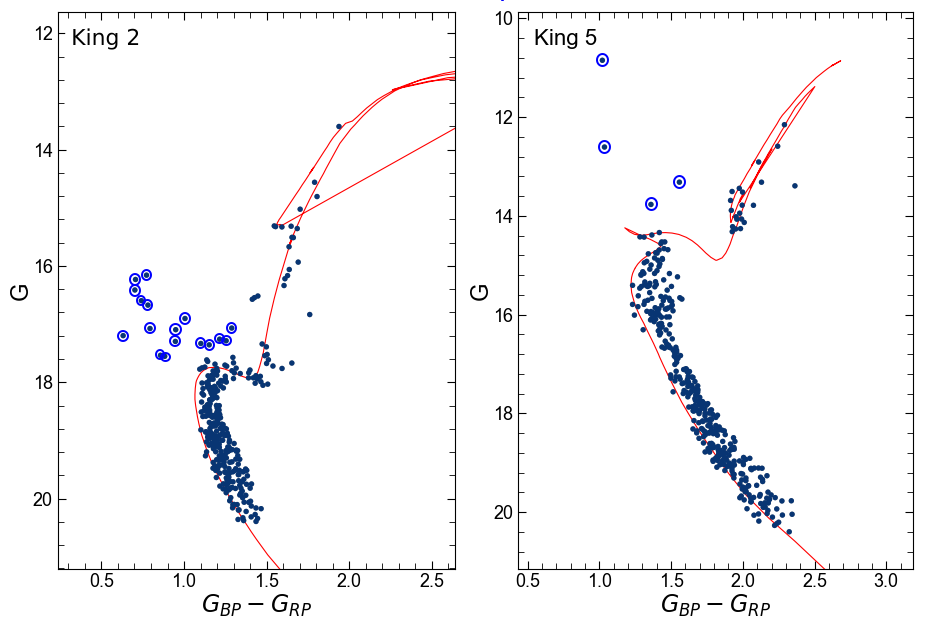}
\caption{CMDs of King 2 and King 5 OCs. The fitted $G$ versus ($G_{\rm BP}-G_{\rm RP}$) isochrones corrected by \cite{Bressan2012} are shown with red lines for different metallicities Z = 0.0088 (King 2) and 0.0109 (King 5) and $t$ (Gyr) = 4.80 $\pm$ 0.30 (King 2) and 1.45 $\pm$ 0.10 (King 5). Blue circle represent BSS candidates.}
\label{fig:figure_06}
\end{figure}

\cite{Rain2021} presented 22 and 4 BSS candidates, respectively, for King 2 and King 5. In this study, we identify potential 17 (King 2) and 4 (King 5) BSSs. These candidate BSSs, marked with a blue circle, are highlighted in Figure \ref{fig:figure_06}. Although the BSS identified in this study are based on previous classifications, further investigation into the connection between BSS and other stellar populations, such as the merging probability in clusters with different stellar densities, is needed.

\begin{table}
\renewcommand{\arraystretch}{1.15}
\centering
\small
\caption{The astrometric and astrophysical parameters for King 2 and King 5, as obtained in this study herewith together with comparisons to other published values.}
\label{tab:table_03}
\begin{tabular}{c|c|c|l}
\hline
Parameter    & King 2  & King 5 & Reference \\
\hline
\hline
$N$	&	340	&	403	&	This study	\\
&	458	&	454	&	\cite{Hunt2023}	\\
\hline
&	-1.407 $\pm$ 0.008&	-0.291 $\pm$ 0.005	&	This study	\\
$\langle {\mu_{\alpha} \cos{\delta} \rangle}$ (mas yr$^{-1}$)	&	-1.420 $\pm$ 0.006	&	-0.304 $\pm$ 0.005	&	\cite{Hunt2023}	\\
&	-1.358 $\pm$ 0.010	&	-0.282 $\pm$ 0.009	&	\cite{Cantat2018}	\\
\hline
&	-0.863 $\pm$ 0.012	&	-1.256 $\pm$ 0.005	&	This study	\\
$\langle {\mu_{\delta}} \rangle$ (mas yr$^{-1}$)&	-0.844 $\pm$ 0.007	&	-1.265 $\pm$ 0.005	&	\cite{Hunt2023}	\\
&	-0.824 $\pm$ 0.013	&	-1.200 $\pm$ 0.009	&	\cite{Cantat2018}	\\
\hline
&	0.180 $\pm$ 0.049	&	0.409 $\pm$ 0.032	&	This study	\\
$\langle \varpi \rangle$ (mas)		&	0.162 $\pm$ 0.005	&	0.406 $\pm$ 0.003	&	\cite{Hunt2023}	\\
&	0.166 $\pm$ 0.010	&	0.367 $\pm$ 0.004	&	\cite{Cantat2018}	\\
 \hline
	&	5556 $\pm$ 1512 	&	2439 $\pm$ 190	&	This study	\\
$d_{\varpi}$ (pc)	&	6173	&	2463	&	\cite{Hunt2023}	\\
	&	6024	&	2725	&	\cite{Cantat2018}	\\
 \hline
$(m-M)_{\rm o}$ (mag)&  14.896 $\pm$ 0.053	&	13.321 $\pm$ 0.038	&	This study	\\
	&	13.687	&	11.994	&	\cite{Hunt2023}	\\
 \hline
&	6586 $\pm$ 164 	&	2220 $\pm$ 40	&	This study	\\
$d_{\rm Photo}$ (pc)		&	5240.828	&	2286.376	&	\cite{Hunt2023}	\\
	&	5377.6	&	2549.9	&	\cite{Cantat2018}	\\
 \hline
$z$	&	0.0088	&	0.0109	&	This study	\\
\hline
$t$ (Gyr)	&	4.8 $\pm$ 0.3	&	1.45 $\pm$ 0.1	&	This study	\\
\hline
$E(B-V)$ (mag) &	 0.306 $\pm$ 0.019	&	0.605 $\pm$ 0.013	&	This study	\\
	&	0.5549	&	0.9999	&	\cite{Hunt2023}	\\
 \hline
$E(G_{\rm BP}-G_{\rm RP})$ (mag)	&	0.431 $\pm$ 0.027	&	0.853 $\pm$ 0.019	&	This study	\\
\hline
$X_{\odot}$ (kpc)	&	-3.006 	&	-1.962	&	This study	\\
	&	-2.778	&	-2.03	&	\cite{Cantat2018}	\\
 \hline
$Y_{\odot}$ (kpc)	&	4.651	&	1.437	&	This study	\\
	&	4.299	&	1.487	&	\cite{Cantat2018}	\\
 \hline
$Z_{\odot}$ (kpc)	&	-0.454 	&	-0.187	&	This study	\\
	&	-0.419	&	-0.188	&	\cite{Cantat2018}	\\
 \hline
$R_{\rm gc}$ (kpc)	&	12.08 	&	10.22 	&	This study	\\
	&	11.92	&	10.476	&	\cite{Cantat2018}	\\
 \hline
\end{tabular}
\end{table}

\section{Luminosity and Mass Functions}\label{sec5}
\begin{figure}
\centering
\includegraphics[width=0.75\linewidth]{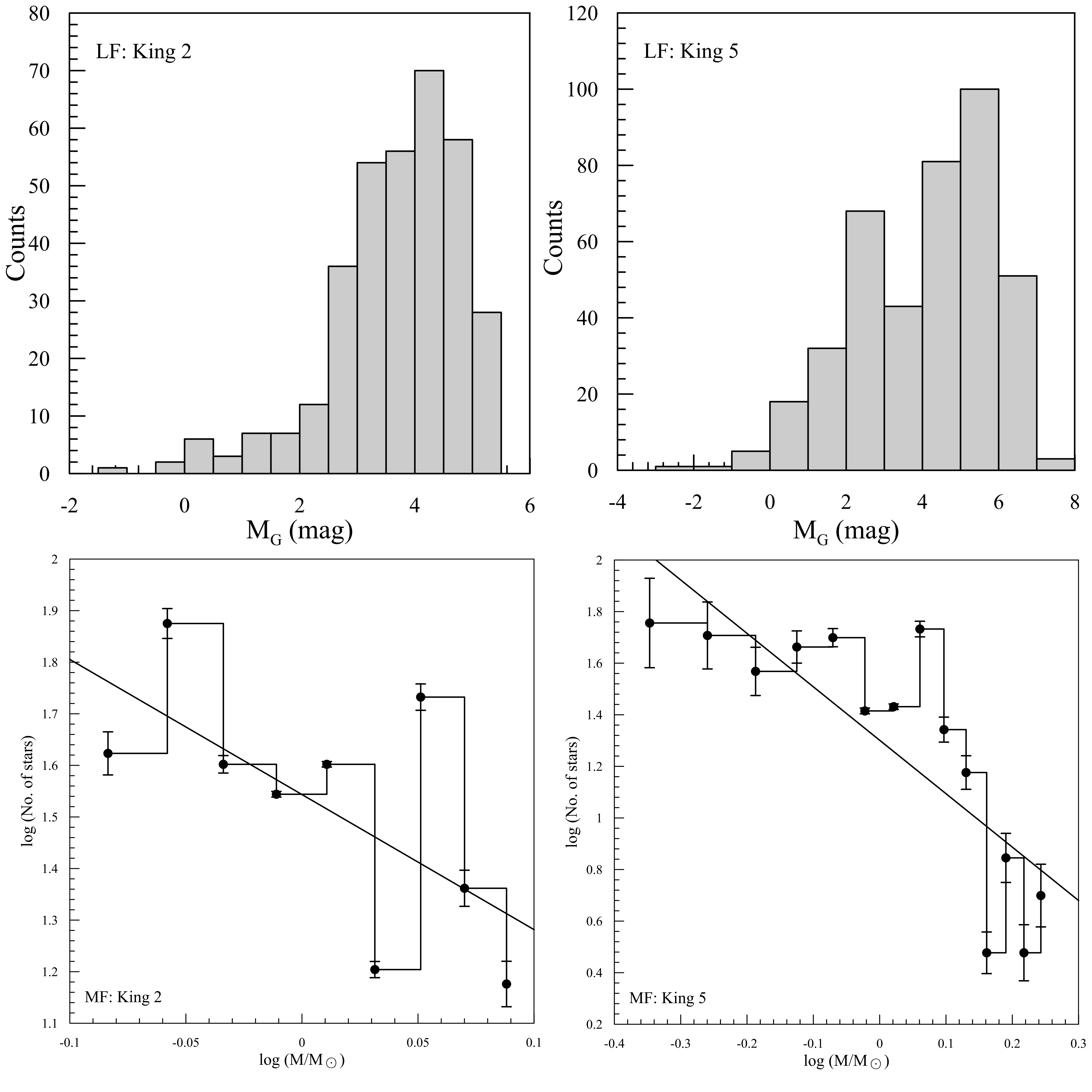}
\caption{TThe true LF of King 2 (\textit{upper left panel}) and King 5 (\textit{lower right panel}), along with the MF of King 2 (\textit{lower left panel}) and King 5 (\textit{lower right panel}), are fitted using the power law from \cite{Salpeter1955} to calculate the slope $\alpha$.}
\label{fig:figure_07}
\end{figure}

After we determined new structural and astrophysical parameters for the King 2 and King 5 open clusters, we used the {\it Gaia} DR3 data and analyzed it with the {\sc ASteCA} packages. We estimate the luminosity functions, (LF) and mass functions (MFs), where the LF is defined as the distribution of members according to different absolute magnitudes; The upper panel of Figure \ref{fig:figure_07} shows the LF member stars, and the corresponding total absolute luminosities $M_{\rm G}$ are listed in Table \ref{tab:table_04}. OCs contain both low and high-mass stars which have the same morphology and formation within the same molecular cloud, this makes them the ideal objects to study the initial mass function IMF \citep{Piatti2002, Yontan2023, Elsanhoury2024, Elsanhoury2025} which is defined as the number density $dN$ of stars scattered over a logarithmic mass scale within a mass range $dM$, centered at mass $M$.

\begin{table}
\centering
\small
\renewcommand{\arraystretch}{1.15}
\caption{Obtained LFs and MFs results of King 2 and King 5 OCs with their $\alpha$'s.}
\label{tab:table_04}
\begin{tabular}{c|c|c}
\hline
Parameter    & King 2 & King 5  \\
\hline
$\langle {M_{\rm G}} \rangle$ (mag) & 3.686 $\pm$ 1.132 & 4.076 $\pm$ 1.820    \\
$a_{\rm 0}$ & 1.18395 $\pm$  0.0071 & 1.71670 $\pm$ 0.0051   \\
$a_{\rm 1}$ & 0.04502 $\pm$ 0.0022  & -0.03456 $\pm$  0.0010    \\
$a_{\rm 2}$ & -0.0204 $\pm$ 0.0001 & -0.01884 $\pm$ 0.0002  \\
$M_{\rm C}$ ($M_{\odot}$) & 356 $\pm$ 19  & 484 $\pm$ 22  \\
$\langle{M_{\rm C}}\rangle$ ($M_{\odot}$) & 1.047  & 1.200  \\
$\alpha$ & -2.62 $\pm$ 0.02  & -2.08 $\pm$ 0.01   \\
\hline
\end{tabular}
\end{table}

Both LF and MF are associated with each other by the well-known Mass-Luminosity Relation (MLR), then by accounting for absolute magnitudes $M_{\rm G}$ (mag) and the masses ($M/M_{\odot}$) combined with adopted isochrones \citep{Bressan2012} on CMDs for estimated ages, distance modulus, and reddening, it is possible to infer the mass-luminosity ratio (MLR) of individual member stars with the second-order polynomial function with coefficients $a_0$, $a_1$, and $a_2$ with their uncertainness as well as the total mass $M_C$($M_{\odot}$) as presented in table \ref{tab:table_04}. The present-day mass function (PDMF) and its dimensionless parameters or slope denoted by ($\alpha$) is given by \cite{Salpeter1955} as -2.35 and indicating the characteristics of massive stars ($> 1 M_{\odot}$). Salpeter's power law suggests that  the star count in each mass range decreases significantly as the mass increases. The values we obtained, $\alpha_{\text{King2}}=-2.62\pm0.02$ and $\alpha_{\text{King5}}=-2.08\pm0.01$, are consistent with the findings of \cite{Salpeter1955}.

The membership probability of a star is calculated based on the statistical assignment, where a star with a membership probability of 50\% is considered to contribute fully to the luminosity function. This approach accurately represents the likelihood of inclusion in our analysis.
The mass of stars in our study is derived using both the MLR and theoretical isochrones, ensuring reliable mass estimates across the observed population. Additionally, we factored out the contamination of field stars by employing the {\sc ASteCA} code, which calculates membership probabilities based on proper motions and color-magnitude data.

The optimal fitting of the data to a straight line was performed using the least-squares method, and the resulting plots for the MF are presented in Figure \ref{fig:figure_07} (lower panel). In this analysis, the $M_{\rm G}$ ranges for MFs calculations contain the stars with; -1.293 $\leq M_{\rm G}$ (mag) $\leq$ 5.496  \& 0.815 $\leq M_{\rm C}$ ($M_{\odot}$) $\leq$ 1.209 (King 2) and -2.470 $\leq M_{\rm G}$ (mag) $ \leq$ 7.367 \& 0.440 $\leq M_{\rm C}$ ($M_{\odot}$) $\leq$ 1.732 (King 5).

\section{Dynamical Evolution Times}\label{sec6}
Dynamically, OCs differ from their compact halo counterparts, such as globular clusters, due to the interactions among their stars leading to exchange energy \citep{Baumgardt2022, Arnold2025}. OCs typically exhibit a looser spatial distribution than globular clusters. Processes such as mass segregation, whereby more massive stars concentrate towards the cluster core compared to fainter ones, have been observed in numerous OCs \citep{Piatti2016, Zeidler2017, Dib2018, Rangwal2019, Bisht2020, Joshi2020}. It has been demonstrated that, as the cluster evolves, its kinetic energy, represented by the velocity distribution of its stars, tends towards a Maxwellian equilibrium \citep{Bisht2019}. The characteristic time required for this dynamical evolution is known as the dynamical relaxation time, denoted by $T_{\rm relax}$ (yr). The mathematical form of this function was established by \citet{SpitzerHart1971}, and it depends on both the number $N$ of member stars and the cluster diameter ($D\approx2r_{\rm lim}$). This function was further explored by \citet{LadaLada2003}:
\begin{equation}
\label{eq:Trelax}
T_{\rm relax} ~=~ \frac{8.9\times10^5~N^{1/2}~ R_{\rm h}^{3/2}}{\sqrt{\langle {M}_{\rm C}\rangle}~log(0.4~N)}
\end{equation}
The half-mass radius ($R_{\rm h}$) values we obtained are 5.22 $\pm$ 0.44 pc and 3.10 $\pm$ 0.57 pc for the respective clusters. The mean mass of the members, $\langle {M}_{\rm C} \rangle$, in solar masses can be determined using the transformation provided by \citep{Sableviciute2006}:
\begin{equation}
R_{\rm h}~=~0.547 \times r_{\rm c} \times \Big(\frac{r_{\rm t}}{r_c}\Big)\textsuperscript{0.486}
\end{equation}
where $r_{\rm c}$ is the core radius and $r_{\rm t}$ is the tidal radius is defined as the distance at which a balance between two gravitational forces acts, one towards the Galactic center and another towards the cluster center, to keep the cluster is bound \citep{Schilbach2019}. The subsequent objective is to estimate the evaporation time ($\tau_{ev} \approx 10^2 T_{\rm relax}$), which characterizes the time required for the ejection of all member stars due to internal stellar interactions \citep{Alvarez2024}. Low-mass stars typically escape the cluster over time, primarily at low velocities through the Lagrange points \citep{Kupper2015}. To maintain the gravitational binding of the cluster, the escape velocity ($V_{\rm esc}$), which is necessary for the rapid expulsion of gas, is computed using the relation \big($V_{\rm esc} = R_{\rm gc} \sqrt{\frac{2GM_C}{3r_{\rm t}^3}}$\big), as described by \citep{Kafle2018, HuntVasiliev2025}, where $R_{\rm gc}$ represents the distance to the Galactic center, $G$ is the gravitational constant, and $M_{\rm C}$ denotes the total mass of the cluster.

In conclusion, the dynamical state of King 2 and King 5 clusters can be described and defined, allowing the dynamical evolution parameter ($\tau=age/T_{\rm relax}$) to be calculated. This demonstrates that the two clusters are dynamically relaxed. All our numerical results for these different dynamical evolution times are listed in Table \ref{tab:table_05}.

\begin{table}
\renewcommand{\arraystretch}{1.15}
\small
\centering
\caption{Dynamical evolution times and escape velocity for King 2 and King 5.}
\label{tab:table_05}
\begin{tabular}{lcccc}
\hline
Cluster    & $T_{\rm relax}$ & $\tau_{\rm ev}$ & $\tau$ &  $V_{\rm esc}$   \\
 & (Myr) & (Myr) & & (km s\textsuperscript{-1}) \\
\hline
King 2 & 59 & 5900 & 81.54 & 71.29 $\pm$ 8.44\\
King 5 & 35 & 3500 & 42.33 &  178.15 $\pm$ 13.35   \\
\hline
\end{tabular}
\end{table}

\section{VEPs and CP}
\label{sec7}
\begin{figure}
\centering
\includegraphics[width=1.10\linewidth]{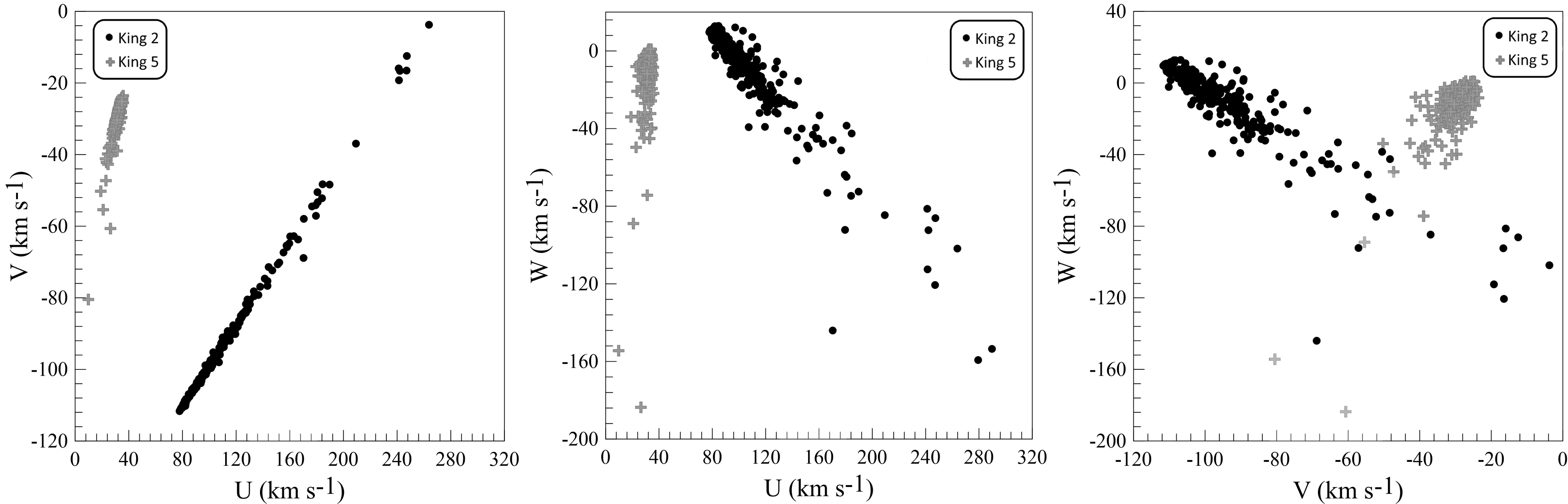}
\caption{The distribution of the spatial space components ($U,~V,~W$) \textbf{of the members} of King 2 and King 5.}
\label{fig:figure_08}
\end{figure}

\begin{figure}
\centering
\includegraphics[width=0.85\linewidth]{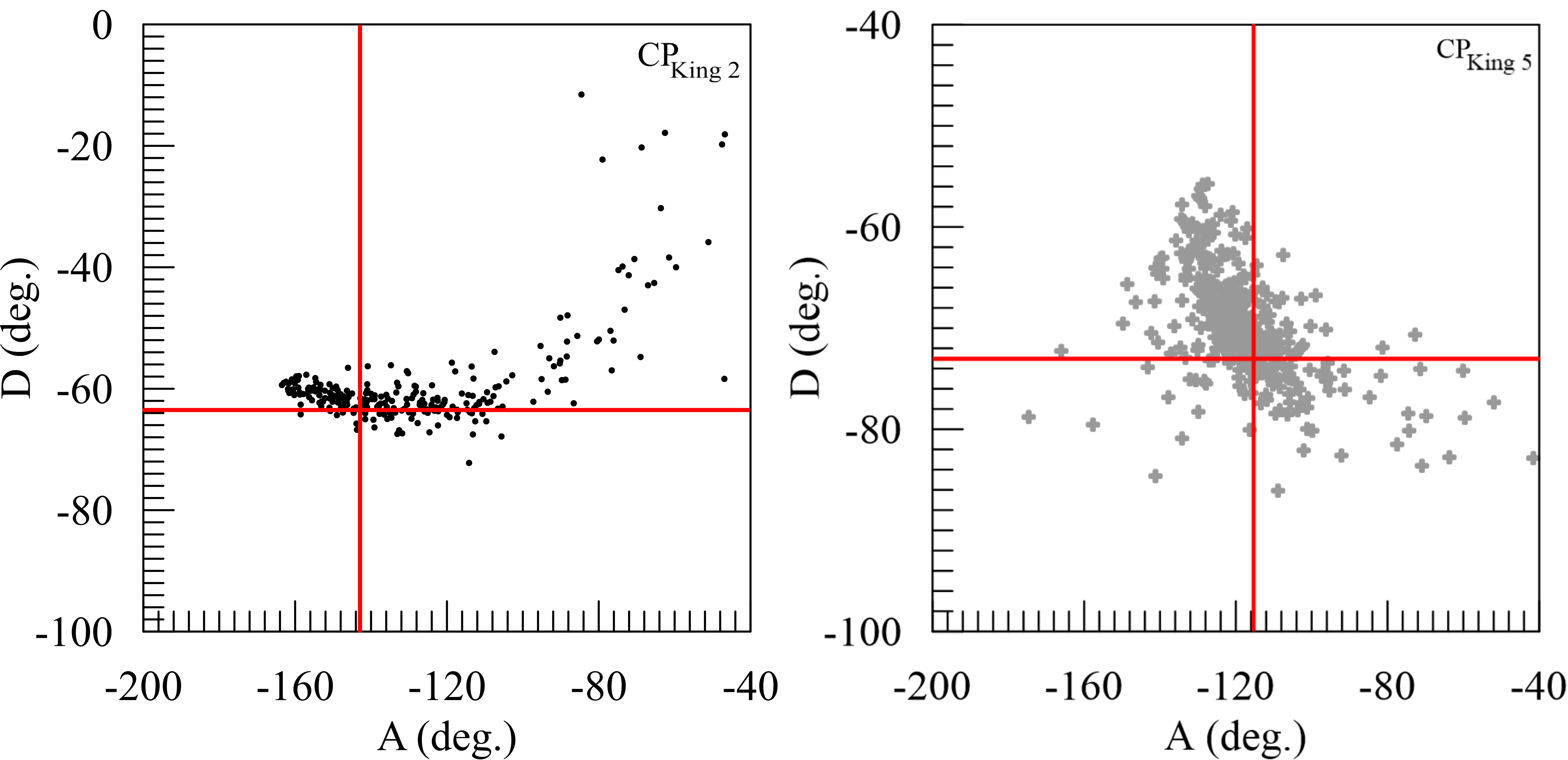}
\caption{Plots of the AD-diagrams for King 2 and King 5. The crosshair refers to the apex position ($A_o,~D_o$) of the members.}
\label{fig:figure_09}
\end{figure}

In order to emphasise the gravitationally bound nature of the stellar groups within a confined volume of space within the Galactic system, we conducted a study of the velocity ellipsoid parameters (VEPs) and kinematics. This was achieved through the utilisation of a computational algorithm applied by \cite{Elsanhoury2022} using member stars with coordinates ($\alpha,~\delta$) located at distances ($d_{\rm i}$; pc) and PMs ($\mu_\alpha\cos\delta,~\mu_\delta$; mas yr$^{-1}$). To estimate the mean radial velocities of King 2 and King 5, we analyzed the most probable members (P $\ge$ 50\% based on $Gaia$ DR3 data. Following the methodology outlined by \cite{Soubiran18}, we applied their equations to derive mean radial velocity values for these clusters. ($V_r$; km s\textsuperscript{-1}) are -136.34 $\pm$ 0.14 (King 2) and -41.94 $\pm$ 0.32 (King 5). In this context, the velocity components ($V_{\rm X},~V_{\rm Y},~V_{\rm Z}$; km s$^{-1}$) along the $x$, $y$, and $z$ axes in a Sun-centered coordinate system, as well as the spatial space velocity components ($U,~V,~W$; km s$^{-1}$) of the member stars on the celestial sphere, along with their Galactic coordinates, can also be calculated \citep[see also,][]{Haroon2024, Elsanhoury2022, Elsanhoury2021}. These values are presented in Table \ref{tab:table_06}. The distribution of these space velocities is shown in Figure \ref{fig:figure_08}. Based on the aforementioned analysis, we can determine the apex position, which represents a point in space, $(A_o,~D_o)$, toward which the member stars are collectively moving. This apex, known as the convergent point (CP), is identified by the intersection of the spatial velocity vectors of individual stars with the celestial sphere. We adopted the AD-diagrams method given by \citep{Chupina2001}, where they use the notion of an individual stellar apex in equatorial coordinates for the members through space velocity vectors ($V_{\rm X},~V_{\rm Y},~V_{\rm Z}$). Therefore, $A_o=\tan^{-1} (\overline V_y/\overline V_x)$ and $D_o=\tan^{-1} (\overline{V_z}/\sqrt{\overline{V_x}+\overline{V_y}})$. Figure \ref{fig:figure_09} gives the apex equatorial coordinates ($A_o,~D_o$) for King 2 and King 5 OCs, and the numerical results devoted with Table \ref{tab:table_06}.

\begin{table}
\renewcommand{\arraystretch}{1.15}
\centering
\small
\caption{Dynamical evolution and kinematic parameters of King 2 and King 5 OCs.}
\label{tab:table_06}
\begin{tabular}{c|c|c}
\hline
Parameter    & King 2 & King 5 \\
\hline
$\overline V_{\rm X}$ (km s$^{-1}$) & 14.61 $\pm$ 3.82 & -5.54 $\pm$ 2.35      \\
$\overline V_{\rm Y}$ (km s$^{-1}$) & -112.58 $\pm$ 10.61 & -11.83 $\pm$ 3.44      \\
$\overline V_{\rm Z}$ (km s$^{-1}$) & -228.52 $\pm$ 15.12 & -43.17 $\pm$ 6.57      \\
\hline
$A_o$ (${\rm ^o}$) & -142.61 $\pm$ 0.08  & -115.10 $\pm$ 0.09\\
$D_o$ (${\rm ^o}$) & -63.58 $\pm$ 0.14  & -73.16 $\pm$ 0.12\\
\hline
$\overline U$ (km s$^{-1}$) & 172.27 $\pm$ 13.13 & 31.61 $\pm$ 5.62      \\
$\overline V$ (km s$^{-1}$) & -58.13 $\pm$ 7.60 & -29.82 $\pm$ 5.46      \\
$\overline W$ (km s$^{-1}$) & -59.23 $\pm$ 7.70 & -12.09 $\pm$ 3.48       \\
\hline
$x_c$ (kpc) & 8.664 $\pm$ 0.09 & 1.118 $\pm$ 0.03      \\
$y_c$ (kpc) &  1.961 $\pm$ 0.04 & 1.272 $\pm$ 0.04      \\
$z_c$ (kpc) & 14.320 $\pm$ 0.12 & 2.224 $\pm$ 0.05      \\
\hline
$S_{\odot}$ (km s$^{-1}$) & 191.22 $\pm$ 13.83 & 45.10 $\pm$ 6.72      \\
($l_A$, $b_A$) (${\rm ^o}$) & (18.65, 18.05) & (43.34, 15.55)      \\
($\alpha_A$, $\delta_A$) (${\rm ^o}$) & (-82.61, 53.58) & (64.91, 73.16)      \\
\hline
\end{tabular}
\end{table}

\subsection{Other Kinematic Structure Parameters}
\textit{(a) The Cluster Center ($x_c,~y_c,~z_c$)}\\
For $N_i$ member stars located at a distance $d_i$(pc) with corresponding ($\alpha_i,~\delta_i$), the cluster center is $x_c$, $y_c$, and $z_c$ \citep{Elsanhoury2022} in units of pc and the obtained results are listed in table \ref{tab:table_06}.\\

\textit{(b) {The Solar Elements}}\\
The components of the Sun's velocities ($U_{\odot}=-\overline{U}$, $V_{\odot}=-\overline{V}$, and $W_{\odot}=-\overline{W}$) are $S_{\odot}$ $=\sqrt{(\overline{U})^2+(\overline{V})^2+(\overline{W})^2}$,  
and the position of the solar apex in Galactic coordinates is ($l_A,~b_A$) and within equatorial coordinates is ($\alpha_A,~\delta_A$) and are drawn like seen in Table \ref{tab:table_06}.\\

\begin{figure}
\centering
\includegraphics[width=0.6\linewidth]{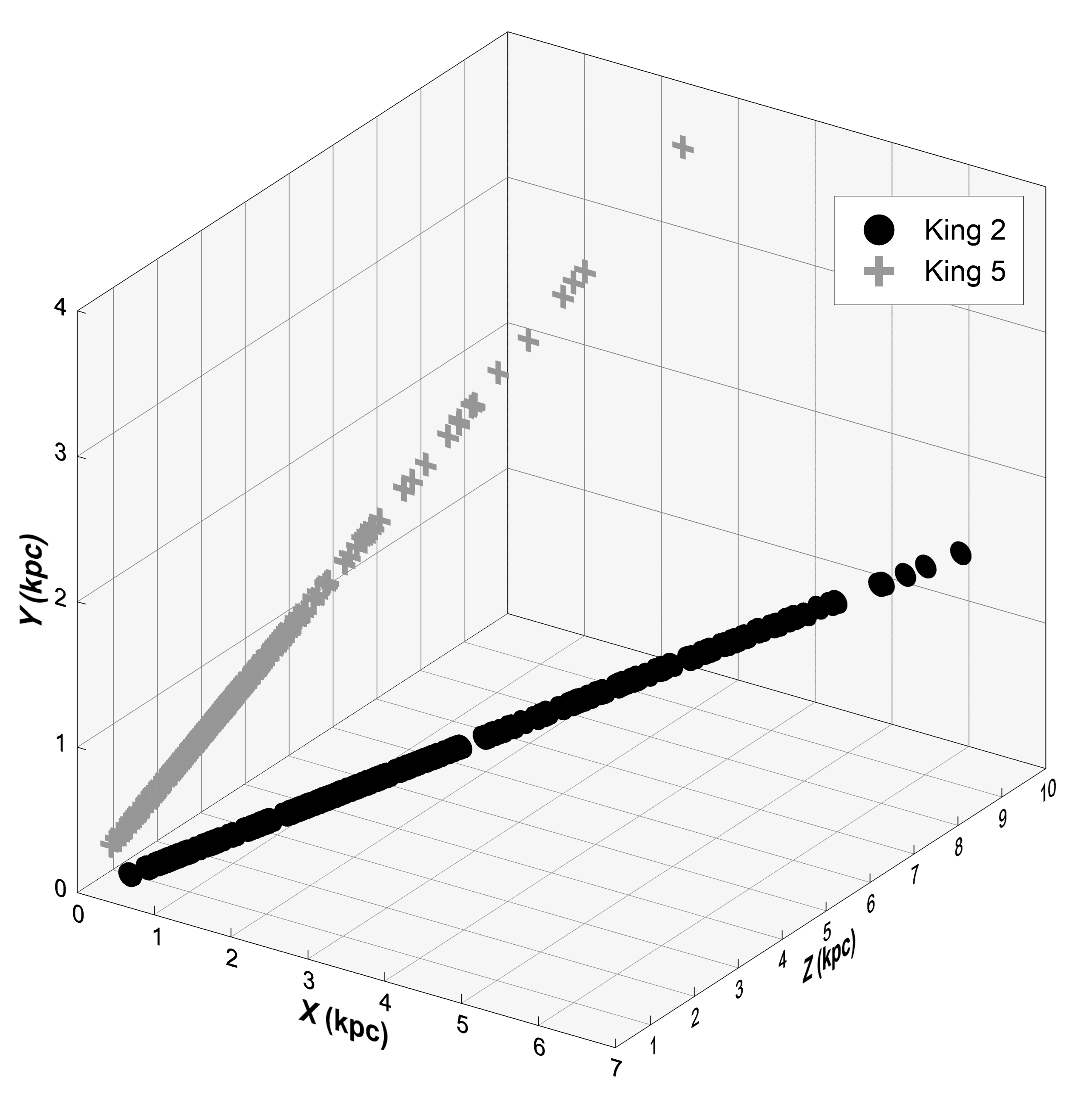}
\caption{The two clusters, designated King 2 and King 5, are illustrated using 3D spatial morphology plots in heliocentric cartesian coordinates ($X,~Y,~Z$; kpc).}
\label{fig:figure_10}
\end{figure}

\textit{(c) King's Morphology with 3D}\\
We analyze the 3D spatial position of member stars in King 2 and King 5 OCs in heliocentric Cartesian coordinates $X$, $Y$, and $Z$ \citep{Elsanhoury2022}. The 3D morphology for these OCs was plotted here in Figure \ref{fig:figure_10}; it is noticeable that the stars are located and expand through separate elongated regions in space. This phenomenon may be regarded as a rapid expulsion and virilization of gas \citep{Pang2021} and accordingly, it can be inferred that the birthplaces of stars are likely located within the same area of the Galactic disc. This suggests that these OCs may have formed from a similar environment or molecular cloud, highlighting their potential shared origins.
\\
\textit{(d) Dynamic Orbit Parameters}\\
The methodology used in this study, which has been successfully applied to both individual stars and other star clusters, allows for a comprehensive analysis of the distribution and dynamics of OCs in the Milky Way \citep{ YontanCanbay2022,TasdemirYontan2023, Yucel2024}. This approach is important for understanding the structure and evolutionary processes of our Galaxy. We determined the Galactic populations of King 2 and King 5 by analyzing the kinematics and dynamics of the clusters' orbits \citep{Dursun2024, Tasdemir2025}.  We also carried out a detailed kinematic analysis, including the calculation of the space velocity components, Galactic orbital parameters and formation radii of the clusters (see Section \ref{sec7}).
\begin{table}
\centering
\footnotesize
\renewcommand{\arraystretch}{1}
\caption{Dynamic orbit parameters of King 2 and King 5 OCs.}
\label{tab:table_07}
\begin{tabular}{lcccccc}
\hline
Cluster & $Z_{\rm max}$ & $R_{\rm a}$ & $R_{\rm p}$ & $R_{\rm m}$ & $e$ & $T_{\rm p}$ \\
&(kpc) &(kpc) & (kpc) &(kpc) & & (Myr) \\
\hline
King 2 & 0.499 $\pm$ 0.246 & 13.076 $\pm$ 1.555 & 7.379 $\pm$ 2.024 & 10.227 $\pm$ 1.789 & 0.279 $\pm$ 0.072& 297 $\pm$ 6 \\
King 5 & 0.177 $\pm$ 0.023 & 10.302 $\pm$ 0.115 & 9.132 $\pm$ 0.232 & ~9.717 $\pm$ 0.174 & 0.060 $\pm$ 0.008 & 277 $\pm$ 5 \\
\hline
\end{tabular}
\end{table}

\begin{figure}
\centering
\includegraphics[width=0.85\linewidth]{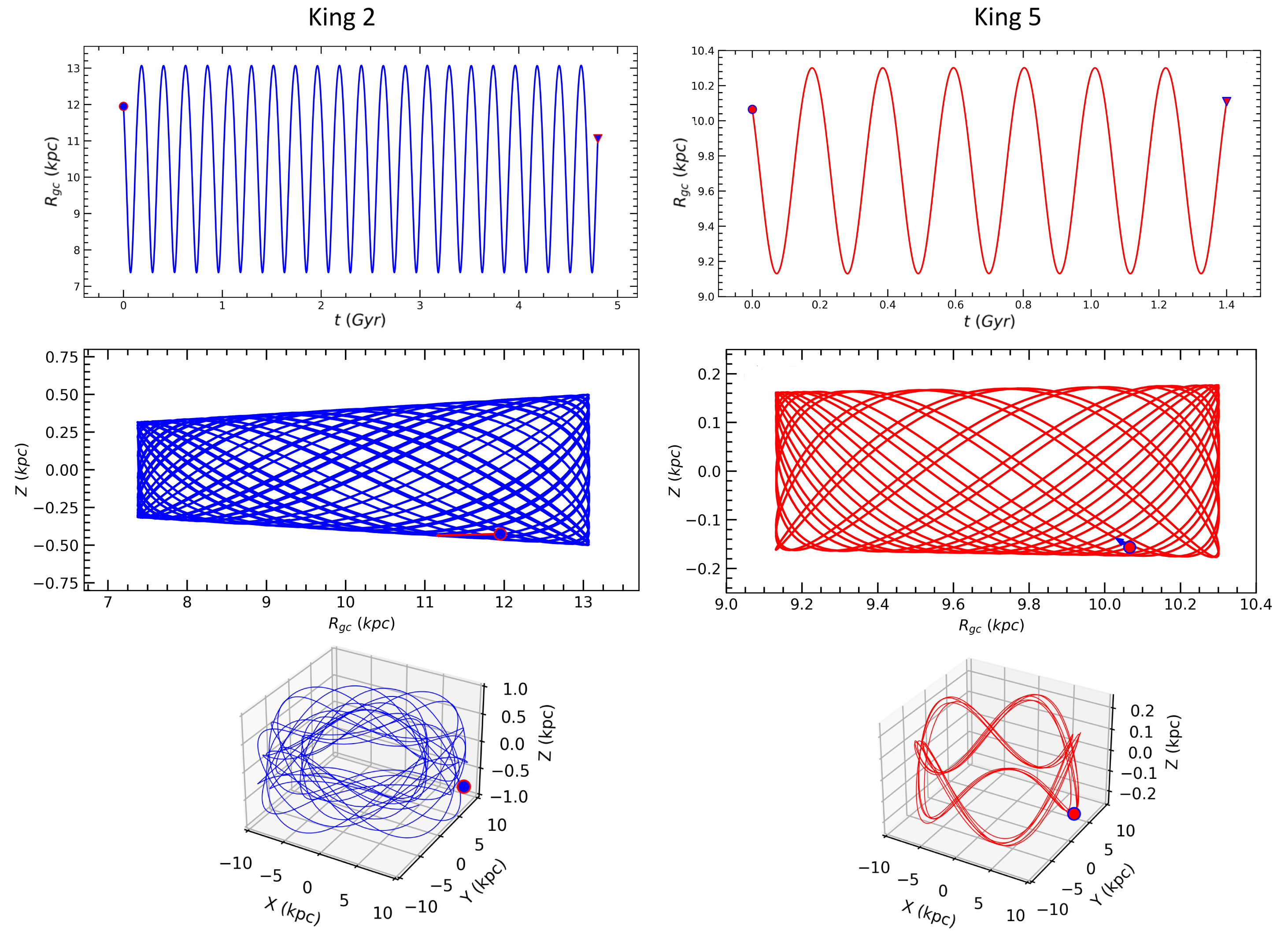}
\caption{The Galactic orbits and birth radii of King 2 and King 5 are illustrated on three different planes: $R_{\rm gc}$ $\times$ $t$  (top), $Z$ × $R_{\rm gc}$ (middle), and $X$ × $Y$ × $Z$ (bottom).}
\label{fig:figure_11}
\end{figure}

In conducting orbital analyses, we employed the {\sc MWPotential2014} model, which is integrated within the {\sc galpy}\footnote{https://galpy.readthedocs.io/en/v1.5.0/} package. This package is a library that facilitates the study of galactic dynamics and was developed by \citet{Bovy2015}. This model incorporates several key parameters, including the Galactocentric distance ($R_{\rm gc}=8.20 \pm 0.10$ kpc) and the orbital velocity of the Sun ($V_{\rm rot}=220$ km s$^{-1}$), \citep{Bovy2015, Eilers2019, Poder2023}. Furthermore, the height of the Sun above the Galactic plane is estimated to be $Z_{\rm 0} = 25 \pm 5$ pc \citep{Juric2008}.

The equatorial coordinates ($\alpha$, $\delta$), distance ($d$), and mean proper motion components ($\mu_\alpha\cos\delta,~\mu_\delta$) derived in Section \ref{sec4}, along with the radial velocity obtained from \cite{Hunt2023}, were used in this analysis. To estimate the current positions of King 2 and King 5, the clusters orbits were traced forward in time using integration steps of 1 Myr up to their respective ages. The orbit integration yielded the following results for King 2: apogalactic distance ($R_{\rm a}= 13.076 \pm 1.555$ kpc), perigalactic distance ($R_{\rm p}= 7.379 \pm 2.024$ kpc), eccentricity ($e = 0.279 \pm  0.063$); for King 5 apogalactic distance ($R_{\rm a}= 10.302 \pm 0.115$ kpc), perigalactic distance ($R_{\rm p}= 9.717 \pm 0.174$ kpc) and eccentricity ($e = 0.06 \pm  0.008$). All orbital parameters are listed in Table \ref{tab:table_07}.

The maximum height reached by King 2 and King 5 above the Galactic plane ($Z_{\rm max}$) is measured as 499 $\pm$  25 pc and 177 $\pm$ 2 pc, respectively. This suggests that both clusters are part of the Milky Way's young stellar disc component \citep{Guctekin2019}. Figure \ref{fig:figure_11} bottom shows the 3D trajectory of the clusters around the Galactic center. As shown, both King 2 and King 5 follow nearly circular orbits while gradually moving away from the Galactic plane during their revolutions. Moreover, Figure \ref{fig:figure_11} (top) illustrates the evolution of the clusters' distances on the $R_{\rm gc} \times t$ plane over time. This representation highlights how uncertainties in the initial conditions influence the orbital paths of the clusters.

\section{Conclusion}
\label{sec8}
In this work, we carried out a comprehensive photometric and kinematical study of the old OCs King 2 and King 5, providing a significant contribution to their characterization by employing an extensive dataset from $Gaia$ DR3, involving a total of 340 member candidates for King 2 and 403 for King 5, with membership probabilities of
$P_{\rm ASteCA}\ge 50\%$. Using the {\sc ASteCA} package, we estimated spatial structures through radial density profiles, verified membership probabilities, and conducted isochrone fitting to determine reddening, distance modulus, metallicity, age, and other astrophysical parameters. Based on the derived luminosity functions and adopted isochrones \citep{Bressan2012}, the masses of King 2 and King 5 were calculated as \(356 \pm 19~M_\odot\) and \(484 \pm 22~M_\odot\), respectively, with mass function slopes (\(\alpha\)) of \(-2.62 \pm 0.02\) and \(-2.08 \pm 0.01\), consistent with \cite{Salpeter1955}. Kinematic analysis indicated the apex positions as \((-142^\circ.61 \pm 0^\circ.08, -63^\circ.58 \pm 0^\circ.13)\) for King 2 and \((-115^\circ.10 \pm 0^\circ.09, -73^\circ.16 \pm 0^\circ.12)\) for King 5. The ages and distances were determined to be \(4.80 \pm 0.30\) Gyr and \(1.45 \pm 0.10\) Gyr, and \(6586 \pm 164\) pc and \(2220 \pm 40\) pc, respectively, with metallicity values of \([Fe/H] = -0.25\) dex (\(z = 0.0088\)) for King 2 and \([Fe/H] = -0.15\) dex (\(z = 0.0109\)) for King 5. Furthermore, we identified 17 and 4 blue straggler stars (BSSs) in King 2 and King 5, respectively, suggesting varied evolutionary processes. Solar elements and dynamical relaxation parameters (\(\tau \gg 1\)) confirmed that both clusters are dynamically relaxed with 3D morphologies elongated within the Galactic disc. Orbital analysis using the {\sc MWPotential2014} model in {\sc galpy} \citep{Bovy2015} revealed apogalactic distances of \(13.076 \pm 1.555\) kpc and \(10.302 \pm 0.115\) kpc, perigalactic distances of \(7.379 \pm 2.024\) kpc and \(9.132 \pm 0.232\) kpc, eccentricities of 0.279 and 0.06, and maximum heights of \(499 \pm 25\) pc and \(177 \pm 2\) pc, firmly placing them within the young stellar disc.

\section*{Acknowledgments}

We express our gratitude to the anonymous referee for their valuable comments and suggestions, which are very helpful in improving our manuscript. This study presents results derived from the European Space Agency (ESA) space mission \emph{Gaia}. The data from \emph{Gaia} are processed by the \emph{Gaia} Data Processing and Analysis Consortium (DPAC). Financial support for DPAC is provided by national institutions, primarily those participating in the \emph{Gaia} Multi-Lateral Agreement (MLA). For additional information, the official \emph{Gaia} mission website can be accessed at \url{https://www.cosmos.esa.int/gaia}, and the \emph{Gaia} archive is available at \url{https://archives.esac.esa.int/gaia}. The author would like to express their gratitude to the Deanship of Scientific Research at Northern Border University, Arar, KSA, for funding this research under project number "NBU-FFR-2025-237-04".

\section*{Data availability statement}
All data that support the findings of this study are included within the article (and any supplementary files).

\section*{Declarations}
The authors declare that they have no known competing financial interests or personal relationships that could have appeared to influence the work reported in this paper.

\section*{Ethical approval}
Not Applicable.

\subsection*{\textbf{ORCiD}}

\textit{A. A. Haroon}: \url{https://orcid.org/0000-0002-8194-5836}

\textit{W. H. Elsanhoury}: \url{https://orcid.org/0000-0002-2298-4026}

\textit{Essam Elkholy}: \url{https://orcid.org/0000-0002-1936-9188}

\textit{A. S. Saad}: \url{https://orcid.org/0000-0002-5145-645X}

\textit{Deniz Cennet \c{C}{\i}nar}: \url{https://orcid.org/0000-0001-7940-3731}


\end{document}